# Multiglass properties and magnetoelectric coupling in uniaxial anisotropic spin-cluster-glass Fe$_2$TiO$_5$

**Shivani Sharma[1], Tathamay Basu[2], Aga Shahee[1], K. Singh[1*], N. P. Lalla[1] and E. V. Sampathkumaran[2]**

[1]*UGC-DAE Consortium for Scientific Research, University Campus, Khandwa Road, Indore - 452001, India*

[2]*Tata Institute of Fundamental Research, Homi Bhabha Road, Colaba, Mumbai- 400005, India*

## Abstract

The compound, Fe$_2$TiO$_5$ (FTO) is a well-known uniaxial anisotropic spin-glass insulator with two successive glassy freezing temperatures i.e. transverse ($T_{TF}$= 9K) and longitudinal ($T_{LF}$= 55 K). In this article, we present the results of measurements of complex dielectric behavior, electric polarization as a function of temperature ($T$), in addition to characterization by magnetic susceptibility and heat-capacity, primarily to explore magnetoelectric (*ME*) coupling and multiglass properties in uniaxial anisotropic spin cluster-glass FTO. The existence of two magnetic transitions is reflected in the isothermal magnetodielectric (*MD*) behavior in the sense that the sign of *MD* is different in the $T$ regime $T<T_{TF}$ and $T>T_{TF}$. The data in addition provide evidence for the glassy dynamics of electric-dipoles; interestingly, this occurs at much higher temperature (~100-150 K) than $T_{LF}$ with high remnant polarization at 10 K (~ 4000μC/m$^2$) attributable to short-range magnetic correlations, thereby offering a route to attain *ME* coupling above 77 K.

**Corresponding Author:**

Dr. Kiran Singh

UGC-DAE Consortium for Scientific Research, University Campus, Khandwa Road, Indore - 452001, India

Phone: +91-731462913; Fax: +91-731462913

E-Mail: kpatyal@gmail.com

## I. INTRODUCTION

In recent literature, investigation of multiferroic properties of materials because of the presence of two or more ferroic properties in one system has proliferated. In multiferroics involving ferroelectricity and magnetism, in general strong magnetoelectric (*ME*) coupling is observed; that is, one can tune magnetization (*M*) and electric-polarization (*P*) or dielectric behavior by electric and magnetic fields (*H*), respectively [1-5]. The tunability of electric-polarization/dielectric with *H* is highly desirable for practical applications. While *ME* coupling is extensively known among magnetically well-ordered materials, such studies have been recently extended to various insulating spin glasses as well [6-10], leading to the recognition for the existence of the phenomenon of 'multiglass' involving glassiness of electric and magnetic states. It is of great interest to explore the nature of *ME* coupling in anisotropic spin-glass systems, in particular to understand whether this behavior is different between longitudinal and transverse freezing regimes.

With this motivation, we have probed the compound $Fe_2TiO_5$ (FTO) for its *ME* studies. This compound, crystallizing in an orthorhombic structure (*Cmcm*), is a well-known uniaxial anisotropic insulating spin-glass [11]. The physics of spin-glasses with single-ion uniaxial anisotropy has been known to be interesting for the following reasons. Depending on the magnetic exchange interaction strength (*J*), uniaxial anisotropy (*D*), external *H* and temperature (*T*), such compounds can exhibit complex magnetic phase diagrams [12, 13]. For a sufficiently small value of $|D|/J$ ratio, two successive freezing transitions i.e. longitudinal ($T_{LF}$) and transverse ($T_{TF}$) can be observed [12-14]. In the case of FTO, two such characteristic temperatures have been known, one at ($T_{LF}$=) 55 K and the other at ($T_{TF}$=) 9 K [11, 15-20]. These two transitions have been presumed to occur by coupling through Dzyaloshinsky-Moriya (*DM*) interaction [18]. Here, on the basis of complex dielectric behavior and electric *P* studies, combined with the results from magnetization and heat-capacity (*C*) measurements, we report that (i) the behavior of *ME* coupling is distinctly different in different temperature regimes i.e. above and below $T_{TF}$ in the magnetically frozen state; (ii) the onset of dielectric anomalies and electric *P* occurs at a temperature (> 100 K) far above $T_{LF}$, as though short-range magnetic correlations are adequate to favor *ME* coupling. This work gains importance considering that there are not many known materials in the literature exhibiting magnetism-induced electric *P* above 77 K and this work demonstrates a route to attain the same; and (iii) a strong frequency ($v$) dependence of complex dielectric behavior near 100-150 K, in addition to frequency dependence in $ac$ susceptibility ($\chi$) at $T_{LF}$, we infer that there is an interesting multiglass dynamics in this material.

## II. EXPERIMENTAL

The polycrystalline sample of FTO has been prepared by a standard solid state reaction method as already reported [21]. The stoichiometric amounts of $Fe_2O_3$ (99.99%) and $TiO_2$ (99.99%) were sintered at 1000°C for 24 hrs with intermediate grindings. Phase-purity was ascertained by x-ray diffraction (XRD, Cu-$K_\alpha$). $DC$ magnetization measurements as a function of $T$ (1.8–300 K) were carried out at 500 Oe for the zero-field-cooled (zfc) and field-cooled (fc) modes using a superconducting quantum interference device (SQUID, Quantum Design) magnetometer. The $ac$ $\chi$ was also measured at different frequencies using the same magnetometer. Complex dielectric behavior (2-300 K) with different frequencies (1- 100 kHz) was obtained during warming (0.5 K/min) using E4980A LCR meter (Agilent technologies). The $T$-dependence of remnant $P$ was obtained from pyroelectric current using Keithley 6517A electrometer. The LCR meter and electrometer were coupled with the (Quantum Design) Physical Properties Measurement System (PPMS). For pyroelectric measurement, the sample was cooled from high temperature to 10 K in the presence of an electric-field (E=740 kV/m) and magnetic field (zero and 140 kOe). The electric field was removed at 10 K and capacitor was short-circuited for 30 minutes to remove the stray charges (if present). The charge $vs$ time behavior was recorded for more than one hour to insure the stability of charge with time before starting temperature dependent measurements. The similar symmetric results were obtained after reversing the direction of poling electric field i. e. E= -740 kV/m. The temperature dependent heat-capacity measurements were performed using PPMS.

## III. RESULTS AND DISCUSSION

Figure 1 shows the Rietveld-refined (FULLPROF Suite) XRD pattern of FTO. As mentioned earlier, this compound crystallizes in (pseudobrookite) orthorhombic structure (*Cmcm*). In this structure, there are four formula units per unit cell. The refined lattice parameters are found to be consistent with the earlier report [21]. On the basis of absence of (112) peak intensity in the powder neutron diffraction data, Atzmony et al. [11] concluded the absence of long-range crystallographic ordering of $Fe^{3+}$ and $Ti^{4+}$ at $4c$ and $8f$ sites, respectively. On the contrary, Guo et al. claims that this system is neither fully ordered nor completely random [21]. Therefore, in the case of partial disorder, one can observe (112) peak with some finite intensity even in the XRD pattern. In the inset of Fig. 1, the presence of (112) peak is demonstrated which clearly confirms the partially ordered nature of the FTO specimen under investigation.

We now present the results of $ac$ and $dc$ $M(T)$ as well as $C(T)$ measurements to bring out the features due to magnetic ordering. Since magnetization behavior has been reported in depth at several places in the past literature [11, 15-20], the discussions of the results are made in brief for the sake of completeness to enable the reader to understand the complex dielectric results. The zfc $dc$ $M(T)$ curve obtained while warming in a field of 500 Oe, shown in figure 2(a) reveals two peaks, one at $T_{LF}$=55 K and the other at $T_{TF}$=9 K in accordance with the earlier results on single crystal [11, 17, 18]. It is worth mentioning that the data on our polycrystalline sample shows more prominent features at $T_{LF}$ and $T_{TF}$ when compared with the past literature on polycrystals, which indicates relatively more crystallographic ordering in our sample. There is a bifurcation of zfc-fc curves near 55 K, which is one of the characteristics of spin-glasses [22]. There are corresponding features in $ac$ $\chi(T)$ data as well (see figure 2b). In addition, real part ($\chi'$) of $ac$ $\chi$ exhibits an observable shift of the cusp at $T_{LF}$ with increasing ν, as observed by Tholence et al. [20] on single crystals. We have also obtained a signature of glassy magnetic state from new results, namely, 'memory' experiments [22, 23]. While zfc curve (called '$M_{zfc\text{-}ref}$') described above was obtained in the sweep mode with negligible waiting time at each measurement temperature, an additional zfc curve was obtained in the following manner: the sample was cooled from 200 K to wait temperature ($T_w$=) 40 K, wait there for time ($t_w$=) 4 hrs and then cooled down to 1.8 K following which the data was collected during warming with the same experimental parameters as mentioned for the $M_{zfc\ ref}$ curve; the curve thus obtained is labeled as $M_{zfc\ w}$. These curves are also shown in bottom inset in figure 2(a). It is clear from the inset that there is a distinct dip at the waiting temperature i.e. at 40 K. This kind of 'local-dip' (known as 'memory effect') was also observed at other waiting temperatures below $T_{LF}$ (not shown here) and the depth of the dip was found to increase with increasing $t_w$. The above-described results firmly establish spin-glass nature of this system [22, 23] below $T_{LF}$.

Further inference for the glassy nature of the magnetic ordering is made from $C(T)$ curve, shown in figure 2c in the form of $C(T)/T$ vs $T$; as expected, the λ-anomaly is absent near 55 K; even near 9 K, though a weak drop is observed around 9 K. A notable feature in the $C(T)/T$ plot is that there is a broad maximum centered around 150 K, which indicates that there is an anomaly around this temperature. The overall behavior of $C(T)$ in this temperature range is dominated by lattice and magnetic contributions. Similar $C(T)$ behavior was also reported for other systems e.g. in CuMn [24] and $Ca_3Co_2O_6$ [25] and attributed to short-range order. At this stage, an iso-structural nonmagnetic reference compound is not available to separate out precisely the lattice and magnetic contributions. One possible explanation can be given in terms of short-range magnetic correlations extending to such a high temperature range. In support of this interpretation, the plot of $\chi^{-1}$ vs $T$ (see inset of figure 2a) is found to be non-linear in a wide $T$-range in the paramagnetic state, in agreement with a previous report [11]. Alternatively, this can be attributed to some other broad transition as that evidenced in this article (see below). In any case, this observation is quite important to one of the main conclusions from magnetic, dielectric and remnant $P$ behavior.

We now address dielectric behavior. Figure 3(a,b) shows the temperature dependent complex dielectric permittivity obtained with different frequencies (1- 100 kHz). It is known that this material is highly insulating at low $T$, which is the prerequisite for the intrinsic nature of dielectric behavior. Consistent with this, the value of $tan\delta$ is also very small. There is no visible sharp change in dielectric constant (ε') as well as in $tan\delta$ at $T_{LF}$ (=55K), though a feeble and gradual fall is observed near $T_{TF}$ (=9K) and can be seen more clearly in figure 3c. Another observation we have made is that there is a strong frequency dependence shoulder in the range of 100 - 150 K (much higher than $T_{LF}$). There is a continuous increase in dielectric constant well above this temperature range. We attribute this to a small increase in electrical conduction known commonly in the literature, rather than attributing it to a global ferroelectric transition setting at such temperatures. The point of central emphasis is that this dipolar cluster glass-like behavior actually sets in well above $T_{LF}$. This is clearly seen in the derivative plots, shown for two frequencies in figure 3c. The peak temperature ($T_m$) of the derivatives are found to obey the power law behaviour i.e. $v = v_0(T_m/T_g - 1)^{zv}$. This is demonstrated in the inset of figure 3c and $T_m$ was obtained from $d(tan\delta)/dT$ plots at different frequencies. The values of fitting parameters $T_g$, $zv$ and $\tau_0$ $(=1/2\pi v_0)$ are 72 K, 4.7 and 1.7 x $10^{-7}$sec, respectively. The observed frequency dependence is typical characteristic of dipolar cluster glass [26]. In order to explore additional characteristics of electric-glassy state, we have performed 'memory' experiments at 30 K after waiting for 15 hrs at 50 kHz (as described above for spin-glass freezing). There is a weak but visible dip in $\Delta\varepsilon'$ at waiting temperature i.e. at 30 K. This is presented in the inset in panel (b) of figure 3; here $\Delta\varepsilon'$ is the difference between $\varepsilon'_{ref}$ and $\varepsilon'_{wait}$ as defined for corresponding magnetization data. The depth of the dip is very small, presumably because such memory effects are broadened due to chemical disorder well-known for this material. The observed frequency dependence may reflect slow electric-dipole dynamics, which are yet to be unraveled. The frequency dependencies in both $ac$ susceptibility and complex dielectric properties are scarce for stoichiometric (that is, undoped) compounds, barring a few exceptions [27, 28]. The demonstrations of distinct memory effect both in magnetic and dielectric studies support multiglass behavior of FTO as was first time observed in the $ME$ multiglass SrTi$_{0.98}$Mn$_{0.02}$O$_3$ [6]. In addition, we have measured $\varepsilon'$ $vs$ $T$ at different magnetic fields up to 140 kOe, but there is no change in these dielectric anomalies (not shown here). As the temperature is lowered, similar to magnetization and heat capacity results, the anomaly at $T_{TF}$ is also observed in dielectric results and can be seen more clearly in figure 3c.

In order to address whether there is any difference in the magnetodielectric behavior across these magnetic transitions, $\varepsilon'$ was measured with 100 kHz as a function of $H$ at various temperatures, the results of which are shown in figure 3d in the form of $MD$ $vs$ $H$ where $MD=[\{\varepsilon'(H)- \varepsilon'(0)\}/\varepsilon'(0)]$. The magnitude of $MD$ is comparable with the reported value for other well-known $ME$ materials i.e. chromite spinel [29]. The results clearly reveal that there are qualitative changes in the behavior as the temperature is lowered. In particular, the sign of $MD$ is different for T<$T_{TF}$ and T>$T_{TF}$ prominently seen above 50 kOe. It is possible that the expected anomaly near $T_{LF}$ (in the plots as a function of $T$) could be broadened due to its sensitivities to some degree of crystallographic disorder (intrinsic to this material due to partial disorder of Fe$^{3+}$ and Ti$^{4+}$ at 8f and 4c sites). The different sign of $MD$ at different $T$-regime is generally scarce, barring exceptions like the case of BiMn$_7$O$_{12}$, hexagonal TbMnO$_3$ and Ca$_3$CoMnO$_6$ [30-32]. The observed *MD vs H* behavior suggests the presence of higher order *ME* coupling effect in this system (mostly quadratic at higher *H*). The sign of *MD* depends on the temperature and magnetic field and can be explained by using the simple phenomenological model given by Katsufuji et al. [33]

$$\varepsilon' = \varepsilon'_0(1 + \alpha <S_i \cdot S_j>_H)$$

where $<S_i \cdot S_j>_H$ is the spin-pair correlation of neighboring spins at a particular applied *H*. This value is negative in case of *AFM* ordering and positive for *FM*/paramagnetic regime. We fitted the *MD* results by using the combination of linear and quadratic terms of *H* i.e. *MD*~ ($\beta_1 H$+ $\beta_2 H^2$) as reported earlier [31, 32]. The coefficient of fitting parameter of quadratic term i.e. $\beta_2$ is plotted as a function of temperature in the inset of panel (d) of figure 3. It can be clearly seen from this inset that the sign of $\beta_2$ is changing at $T_{TF}$ which shows that the sign of *ME* coupling is different at $T<T_{TF}$ and $T>T_{TF}$.

In order to address further *ME* coupling and electric *P*, we have studied remnant *P* behavior, as described in the experimental section. The results are presented in figure 4 (a). The pyroelectric current is also shown in the inset of figure 4 (a). The key finding is that the electric *P* sets in at a very high temperature (>150 K) above $T_{LF}$. The *P* changes slowly with increasing *T* till ~ 120 K, and then decreases sharply, following which, it becomes temperature independent at higher *T*. The sign of *P* is changed by reversing the poling electric field. The value of *P* above 150 K changes continuously, which shows that there is no first order phase transition across this temperature regime. The value of *P* also changes with the application of *H* (for an application of 140 kOe) as shown in the same figure which proves the existence of *ME* coupling. Such *ME* coupling above 77 K is observed in CuO [34] and few other systems also [27, 35]. The remnant *P* in the absence of a magnetic field at 10 K is ~ $4000 \mu C/m^2$ and close to the value reported for $CaBaCo_4O_7$ single crystal [36]. Our results show the qualitative evidence of electric *P* at 740 kV/m poling electric field and quantitatively it can be higher at higher poling field. FTO structure consists of two non-centrosymmetric coordinations i.e. $FeO_6$ and $TiO_6$, and $FeO_6$ octahedron is more distorted than $TiO_6$ octahedra [18]. These results indicate the polar behavior of FTO and its tunability with external *H*.

From the above results, it is thus evident that there exist is a strong *ME* coupling even much above $T_{LF}$. The question arises what is the origin of this feature. We attribute it to the short-range magnetic correlations. In support of this proposal, we refer the quasi-elastic neutron scattering [37] and Mössbauer spectroscopy (*MS*) [19] results which show the presence of short range magnetic correlations well above $T_{LF}$. In conjunction with this the absence of Curie-Weiss behavior in susceptibility (see upper inset of figure 2(a)), $d\chi/dT$ *vs* $T$ and $d^2\chi/dT^2$ *vs* $T$ in figure 4b and its inset, respectively, further establish the existence of short range magnetic correlations well above $T_{LF}$ i.e. around 150 K. Further the presence of short-range correlations up to 150 K can be inferred from figure 4 (b). Around 150 K one can see the change in slope in $d\chi/dT$ *vs* $T$ as well as in $d^2\chi/dT^2$ *vs* $T$ (see inset of figure 4(b)). The broad peak in $C(T)/T$ *vs* $T$ presented in this article is either in support of this short-range magnetic correlations and/or a direct consequence of the local polar behavior in this temperature region. The onset of $P$ (see Fig. 4a) appears at higher temperature than the peak in $d\varepsilon'/dT$ *vs* $T$ (figure 3 (c)). FTO being a disordered material and its frequency dependence dielectric permittivity shows dipolar cluster glass-like behavior and hence $P$ can exist well above the peak in dielectric permittivity due the nucleation of polar-nano regions (PNR). The occurrence of PNR above dielectric peak has been known for relaxor ferroelectrics [26]. Similar to our results, a tail in $P$ above dielectric peak is also observed in Mn doped $SrTiO_3$ [38]. Such a behavior has been reviewed in the article by Cross [39]. In short, we conclude that this compound serves as a rare example among oxides in which *ME* coupling can arise from short-range magnetic correlations, though very recently *ME* coupling and $P$ was reported in the paramagnetic regime in metal-organic framework $[(CH_3)_2NH_2]Mn(HCOO)_3$[40]. Similar findings have been reported in a spin-chain material, $Ca_3Co_2O_6$ [27]. The remnant $P$ is high and comparable with type I multiferroics. This high polarization value could not be solely due to magnetic short range correlations only but can be related to local polar transition also. A high resolution low temperature x-ray diffraction is worthwhile to throw some light on the exact microscopic origin of the observed $P$ results. Finally, it should be noted that the value of $\tan\delta$ is very small up to 150 K i. e. $\tan\delta$=0.04 and the phase angle between current and voltage at 150 K turns out to be about -88° which is very close to the value for an ideal capacitor. Hence, the observed *ME* coupling in this material is its intrinsic properties rather than extrinsic effects like magnetoresistance [41], grain boundary etc.

## IV. CONCLUSION

To summarize, we have reported the evidence for multiglass properties and *ME* coupling in an anisotropic spin-glass, $Fe_2TiO_5$. The sign of magnetodielectric changes as the sample is cooled down from longitudinal freezing temperature across transverse freezing temperature. A key finding is that *ME* coupling is found to set in at a temperature far above longitudinal freezing temperature. Our results demonstrate that short-range magnetic correlations also can trigger *ME* coupling. It should be remarked that such correlations yield spontaneous electric *P* at higher temperatures far above 77 K, thereby offering a route to identify materials for applications involving coupled multiple phenomena. This work gains importance considering current interests to identify *ME* materials above 77 K.

**Figure captions:**

**FIG. 1.** Rietveld refinement of x- ray diffraction pattern of $Fe_2TiO_5$ at room temperature.

**FIG. 2.** For $Fe_2TiO_5$, temperature dependence of (a) *dc* magnetization measured in zfc and fc mode in 500 Oe; the lower inset shows $M_{zfc\ ref}$ and $M_{zfc\ w}$ curves obtained with 4 hrs waiting at 40 K as described in the text; upper inset shows $\chi^{-1}$ *vs T* up to 310 K; (b) real part of *ac* susceptibility measured with $\nu$=0.13 and 13 Hz; inset shows *ac* $\chi$ in an expanded scale near $T_{LF}$; and (c) heat-capacity divided by temperature as a function of temperature; in the inset, the curve is shown in an expanded form in the *T*-range 100-200 K.

**FIG. 3.** Panels (a) and (b) show temperature dependence of real part ($\varepsilon'$) of dielectric constant and *tanδ* for $Fe_2TiO_5$ at different frequencies. The arrows indicate the direction in which curves move with increasing frequency; inset in (b) illustrates the memory effect at 30 K at 50 kHz after waiting at 30 K for 15 hrs. Panel (c) illustrates first derivative of $\varepsilon'$ *vs T* at two frequencies; inset shows power-law dependence of the peak in *d(tanδ)/dT*, as described in the text. Panel (d) shows isothermal magnetodielectric data *MD* [= $(\varepsilon'_H-\varepsilon'_{H=0})/\varepsilon'_{H=0}$] at different temperatures measured at 100 kHz; inset shows the coefficient of quadratic term ($\beta_2$) as a function of temperature (for more details about fitting, see the text).

**FIG. 4.** (a) Temperature dependence of remnant polarization (*P*) at zero and 140 kOe (for details see text), pyroelectric current is also shown in the inset. (b) Derivative of dc $\chi$ measured at 500 Oe; inset shows $d^2\chi/dT^2$ *vs T* behavior.

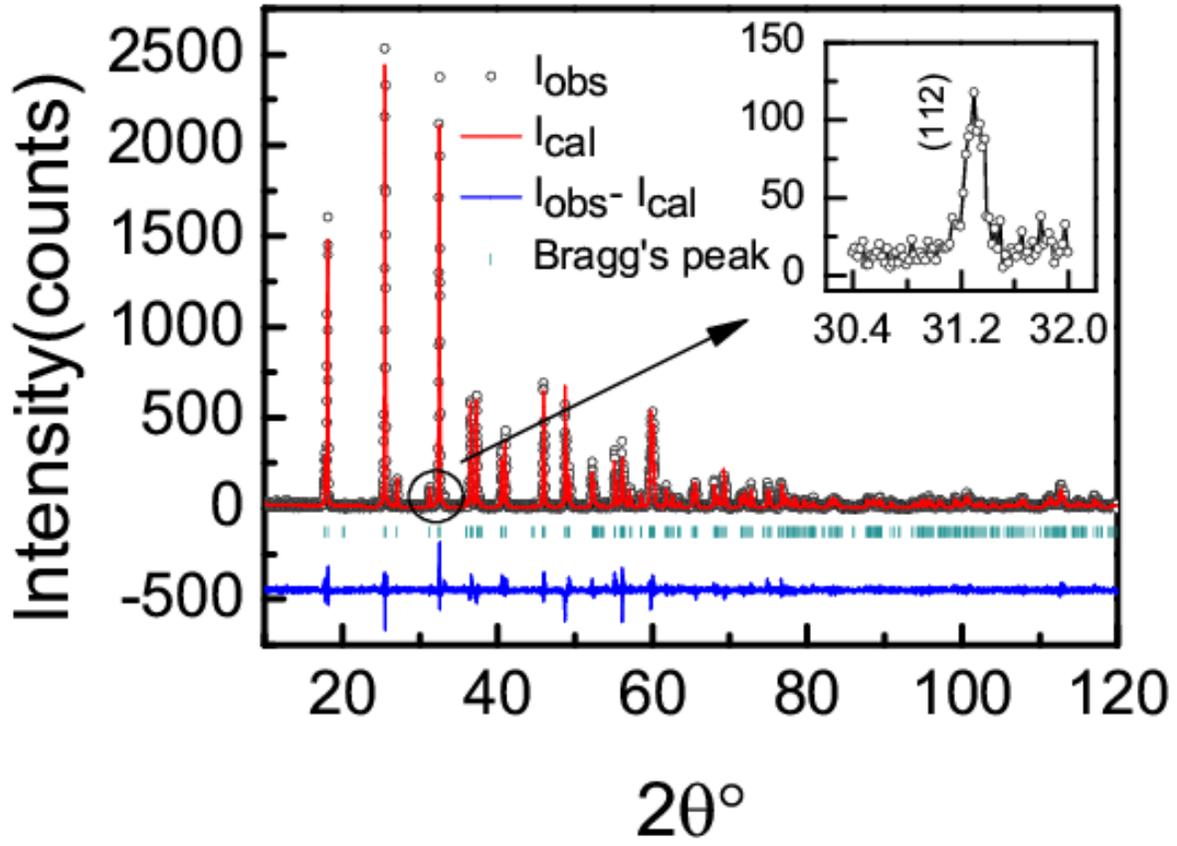

**FIG. 1**

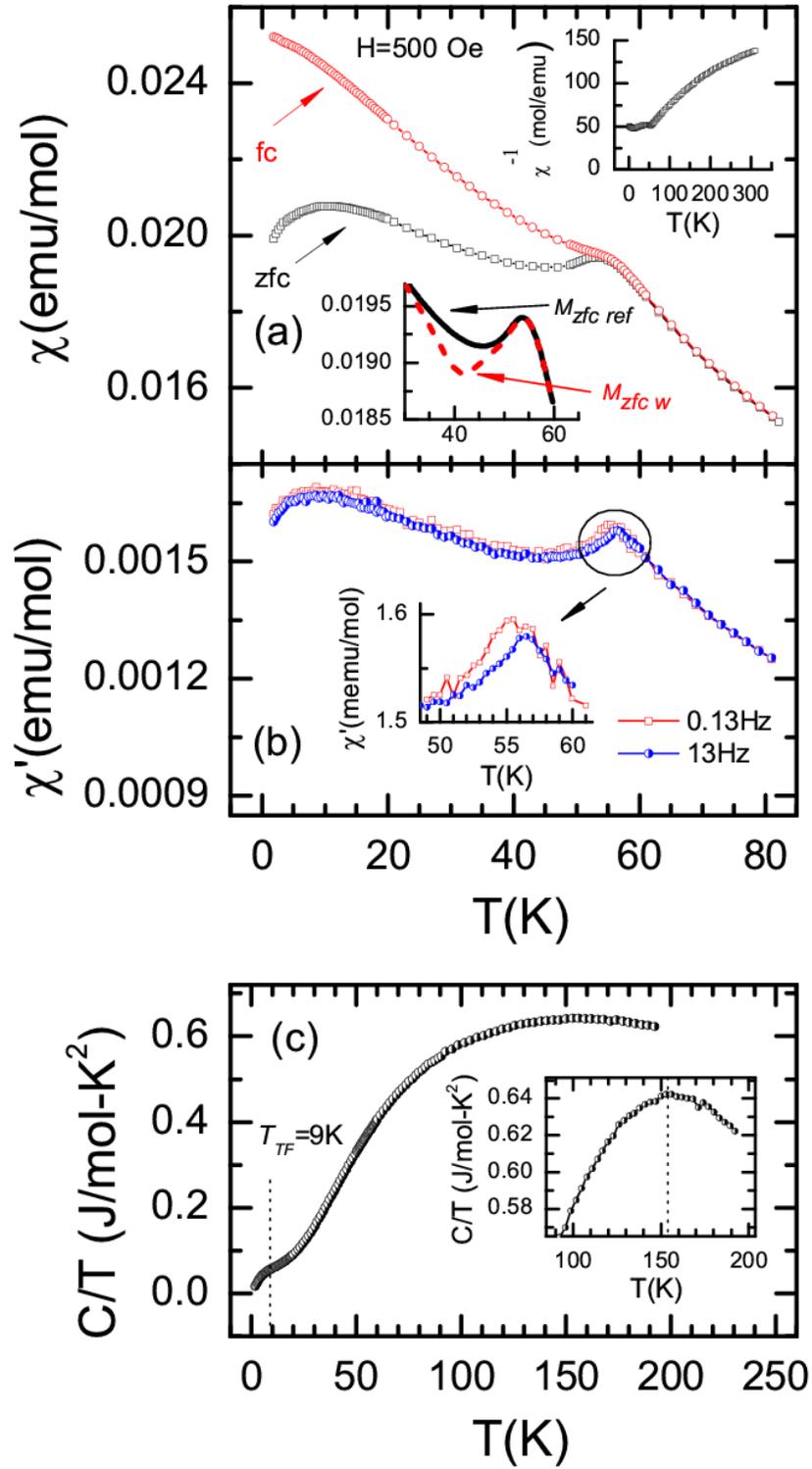

**FIG. 2**

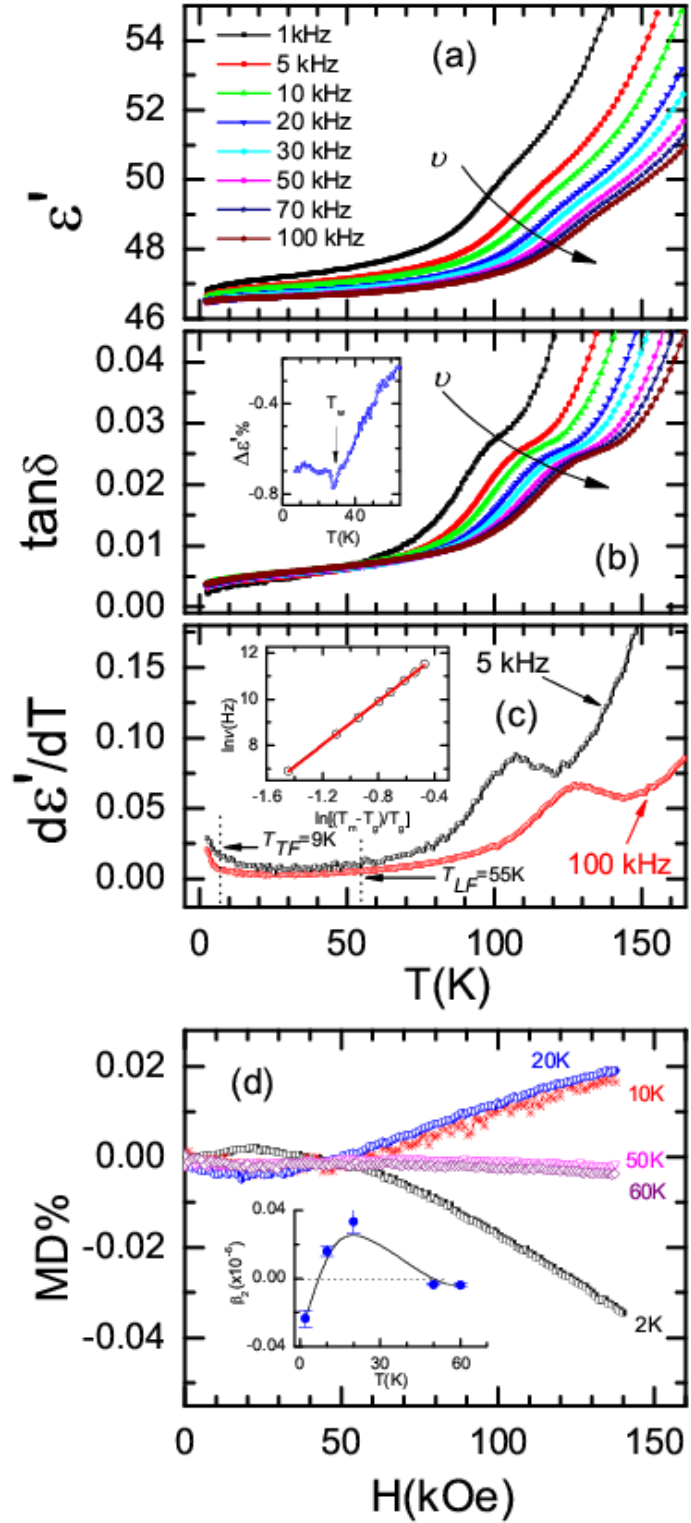

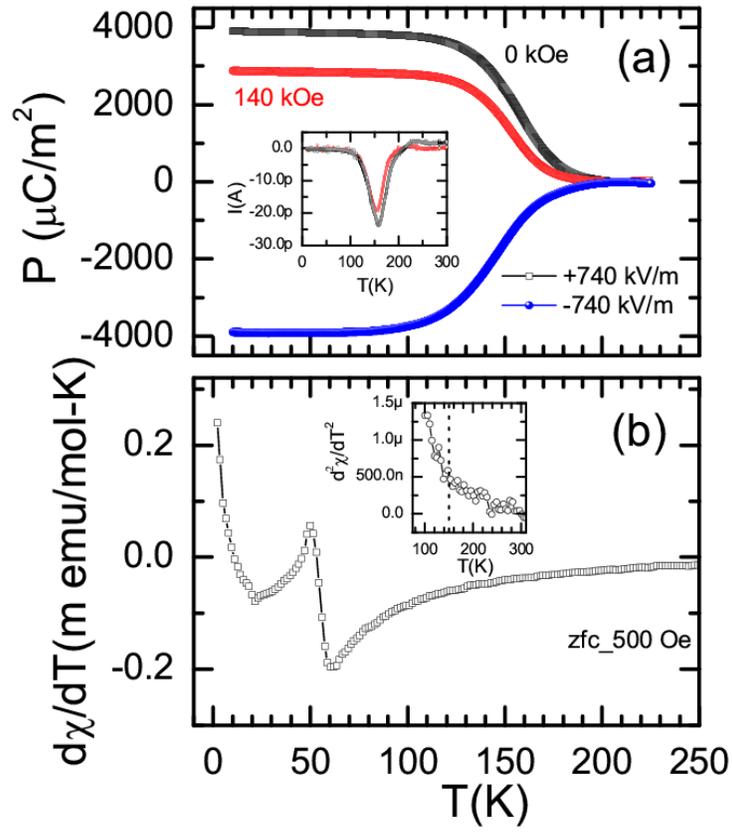

**FIG. 4**